\documentstyle[12pt]{article}
\title{Energy conservation for a radiating charge in classical
electrodynamics} 
\author{Ashok K. Singal\\Astronomy and Astrophysics Division\\Physical Research Laboratory,
Navrangpura\\Ahmedabad - 380 009, India\\Email:asingal@prl.res.in}
\date{}
\begin{document}
\maketitle
\begin{abstract}
\baselineskip=0.9cm
It is shown that the well-known disparity in classical 
electrodynamics between the power radiated in electromagnetic 
fields and the power-loss, as calculated from the radiation 
reaction on a charge undergoing a non-uniform motion, is 
successfully resolved when a proper distinction is made between 
quantities expressed in terms of a ``real time'' and those 
expressed in terms of a retarded time. It is shown that the 
expression for the real-time radiative power loss from a charged 
particle is somewhat different from the familiar Larmor's formula, 
or in a relativistic case, from Li\'{e}nard's formula.
\end{abstract}
\newpage
\baselineskip=0.9cm
\section{Introduction}
One of the most curious and perhaps an equally annoying problem in 
classical electrodynamics is that the power emitted from an 
accelerated charge does not appear to exactly match with the
radiation reaction on the charge. In the standard,
Larmor's radiation formula (generalized to Li\'{e}nard's
formula in the case of a relativistic motion), the radiated
power is directly proportional to the {\em square of acceleration}
of the charged particle. From the energy conservation 
law, it is to be surely expected that the 
power emitted in radiation fields equals the power loss
of the radiating charge. But from the radiation reaction equation,
the power loss of a radiating charge is directly proportional
to its {\em rate of change of acceleration} (see e.g., \cite{1,2,3}).
Although the two formulae do yield the same value
of energy when integrated over a time interval chosen such that 
the scalar product of velocity and acceleration vectors of the 
charge is the same at the beginning as at the end of the interval 
(a periodic motion~!), the two 
calculations do not match when the charge is still undergoing
a non-uniform, non-periodic motion at either end of the time 
interval~\cite{1}. In any case the functional forms of the two formulae
appear totally different. This puzzle has defied a 
satisfactory solution despite the continuous efforts for the last 
80 years or so. It is generally thought that the root-cause of this
problem may lie in the radiation-reaction equation,
whose derivation is considered to be not as rigorous as that of
the formula for power radiated. Some interesting proposals for the
``removal'' of the above discrepancy include 
the ad hoc assumption of an acceleration dependent term either 
in a modified form of the Lorentz-force formula~\cite{4}, or in the 
radiation-reaction equation in the form of a ``bound-energy'' term
for an accelerated charge \cite{5,6} (also called as an 
internal-energy term or simply a ``Schott-term'', based on the first 
such suggestion by Schott \cite{7}), or a somewhat related
proposition \cite{8} that even the {\em proper-mass} of a radiating
charged particle (e.g., of an electron) varies, or even a 
combination of some of these propositions \cite{9,21}. 
Alternately it has been suggested \cite{10} that
there may be some fundamental difference in the electrodynamics of a
continuous charge distribution and of a ``point charge'' (with a 
somewhat different stress-energy tensor for the latter).
It is interesting to note that an understanding
of this anomaly has sometimes been sought beyond the border of the
classical electrodynamics (e.g., in the vacuum-fluctuations of the 
electromagnetic fields in quantum theory \cite{11}). 
Ideally one would expect the classical electrodynamics to be
mathematically consistent {\em within itself}, 
even though it might not be adequate to explain all experimental 
phenomena observed for an elementary charged particle. 
Because of the vastness of literature on this subject, we refer
the reader to a recent review article \cite{12} for further references on 
these and other interesting ideas that have been proposed to remove 
this seemingly inconsistency in classical electrodynamics.

We intend to show here that the difference perceived in the 
two power formulae is merely a reflection of the fact that the 
two are calculated in terms of two different 
time systems. While the radiation reaction formula is expressed 
in terms of quantities being evaluated at the ``real time'' of the 
charged particle, Larmor's radiation formula is 
written actually in terms of quantities evaluated at a retarded time. 
The difference between the two time systems is only $\sim r_{0}/c$ for 
a charged particle of radius $r_{0}$, 
and is as such vanishingly small for a
charge distribution that reduces to a ``point'' in the limit.
But as we will show below, it still gives rise to a finite 
apparent difference, 
independent of $r_{0}$, for the power calculations in the two formulae,
because of the presence of the $1/r_{0}$ term in the self-field
energy of the charge. By bringing out this simple relation between 
these two seemingly contradictory results, we demonstrate
in this way their mutual consistency, 
without invoking any additional hypothesis.
\section{The calculation of self-force}
Poynting's theorem allows us to relate the rate of electromagnetic
energy outflow through the surface boundary of a charge distribution to the
rate of change in the mechanical energy of the enclosed charges. 
Accordingly, in order to
compare the outflow of radiation (as implied by say, Larmor's formula)
from a charged particle that may be undergoing a
non-uniform motion, we need to evaluate the 
self-force ``felt'' by the charge.
We consider a classical, spherical-shell model for 
the charge particle.
We take the motion of the
charged particle to be such that there is
always some inertial frame available in 
which the whole charge is momentarily at rest, i.e., there is no 
differential motion between the various constituents of the charged
particle in its instantaneous rest-frame (a ``rigid-motion'' though 
not a ``rigid-body'' motion \cite{13}). We have to consider the force 
on one part of the charged sphere
due to the fields from its all other parts,
positions of the latter calculated at the retarded times,
and then the force to be integrated over the whole sphere. 
By doing the force calculations in the instantaneous rest-frame
of the charge, we can avoid the computations of magnetic fields.
The electric field at the location of an element of charge $de$ 
due to another element $de'$ is given by \cite{1}, 
\begin{equation}
d{\bf E}=de'\left[\frac{{\bf n}-\mbox{\boldmath $\beta$}}
{r^{2}\gamma ^{2}(1-\mbox{\boldmath $\beta$}.{\bf n})^{3}} + 
\frac{{\bf n}\times\{({\bf n}-\mbox{\boldmath $\beta$})\times
\dot{\mbox{\boldmath $\beta$}}\}}{rc(1-\mbox{\boldmath $\beta$}.
{\bf n})^{3}}\right]_{ret} \;,
\end{equation}
where the quantities on the right hand side are to be evaluated 
at the retarded time. More specifically, {\boldmath $\beta$}$=
{\bf v}/c$, $\dot{\mbox{\boldmath $\beta$}}=\dot{\bf v}/c$, and 
$\gamma=1/\surd(1-\beta ^{2})$ represent respectively the velocity, 
acceleration and the Lorentz
factor of $de'$ at the retarded time, and ${\bf r}=r{\bf n}$ is
the radial vector from the retarded position of $de'$ to the
``present'' position of $de$. While calculating the field arising
from the whole sphere, one has to then also take into
account that the retarded times are different for various
charge elements ($de'$) that belong to different parts of the sphere.
It should be noted that in the case of an accelerated
motion, even if there may be no motion between various
parts of an extended object ($r_{0}>0$) in its instantaneous
rest frame, there will be, in general, a differential
acceleration between various parts of the object, as seen
in other inertial frames. Such a differential acceleration,
which occurs even in the case of a uniformly accelerated
motion (see e.g., ref. \cite{14}), actually represents 
the changing Lorentz contraction of an accelerated object.
A proper account of the work done against the Coulomb forces
of self-repulsion between various parts of a charged body
during its changing Lorentz contraction, is necessary to
get the correct formulae for the total energy in the
Coulomb-fields of such a charge \cite{15}. But such a
differential motion between various parts of the
charge, as seen in a relatively moving frame, can
be ignored as far as its radiation field energy is concerned
(see also the discussion in ref. \cite{7}. 

Actually the mathematical details of the calculations of 
self-force, carried out first by Lorentz~\cite{16} and done later more
meticulously by Schott~\cite{7}, are available in different forms 
in modern text-books \cite{1,2,3,20}. Such calculations are done mostly 
with respect to the instantaneous rest-frame of the charge, but
the results derived can be written in a relativistic
covariant form and then applied in any inertial frame.
We will follow the approach of \cite{2} and \cite{3}, where one 
starts directly from the electric fields (i.e., Eq.~(1) here) 
instead of the potentials as in \cite{1}. It is generally assumed 
in such calculations that the motion of the charged particle
varies slowly so that during the light-travel time
across the particle, any change in its
velocity, acceleration and other higher time derivatives
is relatively small. This is equivalent to the conditions
that $|\dot{\bf v}|\tau \,\ll\,c, \;|\ddot{\bf v}|\tau \,\ll 
\,|\dot{\bf v}|,$ etc., where $\tau = r_{0}/c$.

Under these conditions, one can make a Taylor series expansion
of the retarded quantities in Eq.~(1) around the present time
$t_{0}$; the retarded time being related to the present time by
$t'=t_{0}-r/c$. Strictly speaking, $r$ as well as ${\bf r}$ in Eq.~(1)
do not represent the distance between $de$ and $de'$ at the
present time and need to be measured from the retarded position 
of $de'$ \cite{3}. But it turns out that any net effect of this
subtle difference for a spherically symmetric system
shows up only in terms of order higher than
we are interested in and which become negligible for a small
enough $r_{0}$. Remembering that ${\bf v}(t_{0})=0$ (in the
instantaneous rest frame), we can thus write,
\[
{\bf v}(t')=-\frac{\dot{\bf v}r}{c}+
\frac{\ddot{\bf v}r^{2}}{2c^{2}}+\cdots,
\]
\[
\dot{\bf v}(t')=\dot{\bf v}-\frac{\ddot{\bf v}r}{c}+\cdots,
\]
\[
\left[1-\frac{{\bf v.n}}{c}\right]^{-3}_{ret}=
1-\frac{3\dot{\bf v}.{\bf r}}{c^{2}}
+\frac{3r(\ddot{\bf v}.{\bf r})}{2c^{3}}+\cdots,
\]
all quantities on the right hand side being evaluated at the
present time. Substituting these expressions in Eq.~(1), 
the electric force on $de$ due 
to $de'$ can be written as,
\begin{equation}
d{\bf f}=de de' \left\{ \left[\frac{{\bf r}}{r^{3}}
+\frac{\dot{\bf v}}{rc^{2}}
-\frac{\ddot{\bf v}}{2c^{3}}
-\frac{3{\bf r}(\dot{\bf v}.{\bf r})}{r^{3}c^{2}}
+\frac{3{\bf r}(\ddot{\bf v}.{\bf r})}{2r^{2}c^{3}}\right]+
\left[\frac{{\bf r}(\dot{\bf v}.{\bf r})}{r^{3}c^{2}}
-\frac{\dot{\bf v}}{rc^{2}}
-\frac{{\bf r}(\ddot{\bf v}.{\bf r})}{r^{2}c^{3}}+
\frac{\ddot{\bf v}}{c^{3}}\right]\right\}.
\end{equation}
All other higher order terms will give negligible contribution
for a small $r/c$.
The total self-force on the spherical charge $e$ can be found by 
integrating the above expression over both $de$ and $de'$.
It should be noted here that the net force contribution
from two charge elements, when taken in pairs,
does not cancel, i.e. the force on $de'$ due to $de$
is not equal and opposite to that on $de$ due to $de'$.
But one can exploit the spherical symmetry of the charge distribution 
to cancel some terms upon integration. This is due to the
fact that the average of 
${\bf r}(\dot{{\bf v}}.{\bf r})$ for all possible angles 
over a spherically symmetric region 
is $r^{2}\dot{\bf v}/3$, while ${\bf r}$ averages to zero \cite{2}.

Now we would like to emphasize two interesting points that
do not seem to have been highlighted in such calculations before.\\
(i) The velocity-fields (terms contained within the first 
square bracket on the right hand 
side in Eq.~(2)) contribute nothing to the electromagnetic 
self-force of a charge.
The contributions of the various terms get cancelled when 
integrated over the whole spherical distribution.  
Since the self-force from the 
velocity-fields is zero in the instantaneous rest frame of the 
charge, from the Lorentz transformation it is also zero
in other inertial frames of reference.
Of course a nil self-force from velocity fields is an obvious 
expectation in the case of a charge moving with a uniform velocity,
where this is the only term in the electromagnetic 
field of the charge. But the contribution of velocity fields 
to the net self-force turns out to
be nil even in the case of an accelerated charge, when the  
retarded time values are used in the field calculations for a
spherical charge distribution.\\
(ii) The only non-zero terms that remain in the net self-force are,
\begin{equation}
{\bf f}=e^{2} \left[-\frac{2\dot{\bf v}}{3r_{0}c^{2}}
+\frac{2\ddot{\bf v}}{3c^{3}}\right],
\end{equation}
which can be written as,
\[
{\bf f}=-\frac{2e^{2}}{3r_{0}c^{2}}[\dot{\bf v}-\ddot{\bf v}\tau],
\]
or
\begin{equation}
{\bf f}=-\frac{2e^{2}}{3r_{0}c^{2}}\dot{\bf v}_{ret},
\end{equation}
where $\dot{\bf v}_{ret}=\dot{\bf v}-\ddot{\bf v}\tau$ represents
the acceleration that the charge had at a time $t_{0}-\tau$
(to a first order in $\tau$). 
Thus the ``present'' value of the net self-force 
(including the radiation-reaction drag force) 
``felt'' by a spherical-shell charge, 
at any instant, turns out to be directly proportional to the 
acceleration that the charge (as a whole) was undergoing at a 
time $\tau$ earlier. 
\section{The rate of work done against self-force}
To calculate the instantaneous rate of work being done 
against the self-force of a moving charge in an inertial frame,
one has to take the scalar product of 
the {\em present value} of the self-force 
(which incidentally is proportional to a {\em retarded value} 
of the acceleration, Eq.~(4))
and the {\em present} velocity, ${\bf v}$, of the charge,
both measured in that frame. One can derive the relevant formulae 
with respect to a frame in which
the charge has a non-relativistic motion, and then
use the condition of relativistic covariance to get the more
general formulae valid for any inertial frame. For a non-relativistic
motion, the expression for force can be used directly from that 
in the instantaneous rest frame (Eq.~(4) above).

Accordingly, the rate of work done against self-fields of an accelerated
charge is given by,
\begin{equation}
\frac{d{\cal E}}{dt}=\frac{2e^{2}}{3r_{0}c^{2}}(\dot{\bf v}_{ret}
{\bf .v})\;,
\end{equation}
which, equivalently, in terms of the
``present'' time (real-time) quantities is written as,
\begin{equation}
\frac{d{\cal E}}{dt}=\frac{2e^{2}}{3r_{0}c^{2}}(\dot{\bf v}
{\bf .v}-\ddot{\bf v}{\bf .v}\frac{r_{0}}{c})\;.
\end{equation}
The more familiar form for this expression is \cite{1,2,3},
\begin{equation}
\frac{d{\cal E}}{dt}= \frac{4U_{0}}{3c^{2}}\dot{\bf v}{\bf .v}
-\frac{2e^{2}}{3c^{3}}\ddot{\bf v}{\bf .v}\;,
\end{equation}
where $U_{0}=e^{2}/2r_{0}$ represents the electromagnetic 
self-energy in Coulomb fields of a
spherical-shell charge that is permanently
stationary in an inertial frame. The first term on the right
hand side in Eq.~(7) represents the change in the self-Coulomb field 
energy of the charge as its velocity changes. This term when combined
with the additional work done during a changing Lorentz contraction 
(not included in Eq.~(7)) against the Coulomb self-repulsion 
force of the charged particle,
on integration leads to the correct expression for energy in fields 
of a uniformly moving charged particle~\cite{15}. Thus it is only the 
second term on the right hand side of Eq.~(7) 
that represents the ``excess'' power going
into the electromagnetic fields of a charge with a non-uniform
motion. It has always seemed enigmatic that if Larmor's
formula indeed represents the instantaneous rate of power 
loss for an accelerated charge, why the above term contains 
$-\ddot{\bf v}{\bf .v}$ instead of $\dot{\bf v}^{2}$.

Now a point that does not seem to have been realized before is 
that if in Eq.~(5) we kept the acceleration in terms of its value 
at the 
retarded time and instead express the velocity also in terms of its 
value at the retarded time ($t_{0}-\tau$), then for 
${\bf v}={\bf v}_{ret}+\dot{\bf v}_{ret} \tau$ (to the required
order in $\tau$) we get, 
\begin{equation}
\frac{d{\cal E}}{dt}=\frac{2e^{2}}{3r_{0}c^{2}}(\dot{\bf v}_{ret}
{\bf .v}_{ret} +\dot{\bf v}_{ret}^{2}\tau)\;,
\end{equation}
or
\begin{equation}
\frac{d{\cal E}}{dt}=\left[\frac{2e^{2}}{3r_{0}c^{2}} \dot{\bf v}
{\bf .v}+\frac{2e^{2}}{3c^{3}}\dot{\bf v}^{2}\right]_{ret} \;.
\end{equation}

Thus if we examine the rate at which energy is pouring into the 
electromagnetic fields of a charged particle at some given instant, 
then Eq.~(7) gives the rate in terms of the real-time values of 
quantities specifying the motion of the charge at that particular 
instant. On the other hand, Eq.~(9)
yields the familiar Larmor's radiation formula
(second term on the right hand side), but at a cost that 
a real-time rate of field energy outflow from the
charge distribution is expressed in terms of quantities 
specifying the motion of the charge at a retarded time, 
and not in terms of their values at the present time. 
The time difference $\tau$ may be exceedingly small (infinitesimal 
in the limit $r_{0}\rightarrow 0$), but still its effect in the energy 
calculations is finite because of the presence of the $1/r_{0}$ term for 
energy in self-fields outside the sphere of radius $r_{0}$. 
Now $\tau$ is actually the time taken
for a signal to reach from the centre of the sphere 
to the points at its surface. Essentially it implies that if the
electromagnetic energy outflow through the boundary of the spherical
charge distribution (as inferred from the rate of 
work done against the self-force 
of the charge distribution) is expressed in terms of the 
quantities describing the {\em retarded motion} of an
equivalent ``point'' charge at the center of the sphere,
we obtain the familiar Larmor's formula. We may also point out
that even in the standard text-book statement of
Larmor's formula for radiation from a non-relativistically moving
point charge (see e.g. ref.~\cite{1}), the rate of energy flow at a 
time $t$ through a spherical surface of radius $r_{0}$ 
is written always in terms of the retarded value of the acceleration
(at time $t-r_{0}/c$) of the point charge, the expression being 
exactly equal to the second term on the right hand side of Eq. (9).  

This not only resolves the apparent 
``discrepancy'' in the two power formulae, but also shows an intimate 
relation between the energy in the radiation fields and that in the 
Coulomb fields. In particular, the factor of 
4/3 in the electromagnetic inertial mass ($=4U_{0}/3c^{2}$) 
of a spherical charge in classical electrodynamics
is intimately connected with the factor of 2/3 found both in 
Larmor's formula and in the radiation-reaction formula. For long, 
this ``mysterious'' factor of 4/3 has been considered to be
undesirable and modifications in classical electromagnetic theory
have been suggested to get rid of this factor (see e.g. references
cited in \cite{12,15}). If one does adopt such a modification, then 
one can not see the relation between Larmor's formula and the 
radiation-reaction equation. Moreover, as explicitly shown in 
\cite{15}, the factor of 4/3 in
the electromagnetic inertial mass has a natural explanation 
in the conventional electromagnetic theory 
when a full account is taken of all the work done by 
the electromagnetic forces during the process of attaining 
such a charge distribution.

We can generalize the formula for radiative losses, as given by
the second term on the right hand side of Eq.~(7), to a relativistic 
case, by using the 
condition of relativistic covariance (see e.g. ref. \cite{2}). The 
excess rate of change in the electromagnetic field energy 
(over and above that needed for a motion of the charge with the 
``present'' velocity) can in this way be shown to be,
\begin{equation}
P=-\frac{2e^{2}}{3c^{3}}\gamma ^{4}\left[\ddot{\bf v}{\bf .v}+
3\gamma ^{2}\frac{(\dot{\bf v}{\bf .v})^{2}}{c^{2}}\right]\;.
\end{equation}
This should be contrasted with the rate implied by Li\'{e}nard's
formula, where the instantaneous power radiated is
supposed to be \cite{1},
\begin{equation}
P=\frac{2e^{2}}{3c^{3}}\gamma ^{4}\left[\dot{\bf v}^{2}+
\gamma ^{2}\frac{(\dot{\bf v}{\bf .v})^{2}}{c^{2}}\right]\;.
\end{equation}
Actually the Eqs.~(10) and (11) yield identical results for the
instantaneous power rate in the case of a circular motion. This is
to be expected since $\dot{\bf v}{\bf .v}=\dot{\bf v}_{ret}
{\bf .v}_{ret}=0$ in this case. Even for other periodic 
motions (e.g., a one--dimensional harmonic oscillation of a 
charge, say, within a dipole antenna) the average radiation rate 
over a full cycle is the same from both formulae. But 
for a more general case, the two formulae may not yield identical
results. Specifically, in the case of a uniformly accelerated
charge (i.e., for $\ddot{\bf v}+3\gamma ^{2}\dot{\bf v}
(\dot{\bf v}{\bf .v})/c^{2}=0$, \cite{5}), while Eq.~(10) yields a nil 
rate, Eq.~(11) implies a constant finite rate of radiation throughout.
As is well known, the self-Coulomb field
energy of a charge moving with a uniform velocity is different
for different values of the velocity (see e.g., ref \cite{15}). Now
for a uniform acceleration case the sum of both terms 
on the right hand side of Eq.~(9), as evaluated at the retarded time, 
represents just the rate of increase in self-Coulomb field 
energy of the charge as measured at the ``present'' time
($[{\bf v}.\dot{\bf v}+\dot{\bf v}^{2}r_{0}/c]_{ret}=
{\bf v}.\dot{\bf v}$, for $\dot{\bf v}_{ret}=\dot{\bf v}$).
Thus the Larmor term, as calculated
at the retarded time, in this case just compensates for the fact that 
the rate of change in the self-Coulomb field energy of a charge, as 
indicated by its velocity and acceleration values at the retarded 
time, is different from that presently required. In fact from 
detailed analytical calculations it indeed turns
out \cite{17,18} that for a uniformly accelerated charge, total 
energy in the fields (i.e., including both the velocity
and acceleration field terms from Eq.~(1)) at any 
time $t_{0}$ is just equal to the self-energy of a 
charge moving uniformly with a velocity equal to the 
instantaneous ``present'' 
velocity of the accelerated charge (even though the
detailed field configurations may differ in the two cases).
There is no other excess (radiation!) energy in fields in
this case. It is only in the case of a changing acceleration
that the present rate of work done against self-fields (proportional 
to a {\em retarded} value of acceleration because of the 
time-retarded effects of the self-interaction) is different
from the rate demanded by its changing self-Coulomb field energy
(as calculated from the {\em present} value of acceleration), this
difference ultimately representing the radiated power. 
Since for a uniform acceleration case the retarded value 
of acceleration is the same as its present value, 
there is no radiation. It is now clear that the radiation losses 
for an electric charge do not merely result from its accelerated
motion, but are instead a consequence of {\em a change in the
acceleration}. Moreover it is also obvious that
Larmor's formula or its relativistic generalization,  
Li\'{e}nard's formula, do not unequivocally represent the
radiated power losses from an accelerated charge, in the
most general case.

Dirac~\cite{19} gave an alternative derivation of the radiation reaction 
equation that involved both retarded and advanced potentials,
which is sometimes open to criticism based on the causality
arguments. Moreover, from Dirac's treatment, where the self-energy
term was eliminated altogether, one can not see the relation between
the radiative losses inferred from the radiation reaction 
equation and those implied by 
Larmor's formula, without some ad hoc assumption of an acceleration
dependent ``bound-energy'' term, as mentioned earlier. 
In that sense, the approach of Lorentz and Schott for
calculating the radiation reaction from the self-force of
a charge distribution is preferable to 
that of Dirac. The question of the rest-mass of an {\em actual}
elementary charged particle lies beyond the scope of classical 
electrodynamics, but the divergence of the self-energy of a
``point'' charge (i.e., when $r_{0}\rightarrow 0$) in classical 
electrodynamics could (at least in principle) be taken care of by 
invoking the presence of binding forces 
(needed in any case for the stability of the
classical charged particle) that are
equal and opposite to the electromagnetic forces of self-repulsion 
and assuming further that the observed rest-mass of a
charged particle is a combined effect of the two. But as we
have shown the concept of self-field energy can not be altogether
divorced from that of radiation in a consistent 
classical electrodynamical approach.

{}
\end{document}